\documentclass{ws-procs10x7}
\usepackage{euscript}
\usepackage{amsmath}
\usepackage{amssymb}
\usepackage{epsfig}
\usepackage{cite}
\usepackage{color}

\def\mathswitchr#1{\relax\ifmmode{\mathrm{#1}}\else$\mathrm{#1}$\fi}

\newcommand {\pslash}{\hbox{$\not\hbox{\kern-2.3pt $p$}$}}

\def\rQCED{{\rm QCED}}

\def\alf1{ {\alpha\over\pi} }

\begin{document}
\title{QED$\otimes$QCD Exponentiation: Shower/ME Matching and IR-Improved DGLAP Theory at the LHC}
\author{B.F.L. Ward and S.A. Yost}

\address{Department of Physics, Baylor University, Waco, TX, USA}



\twocolumn[\maketitle\abstract{
We discuss the elements of QED$\otimes$QCD
exponentiation and its interplay with shower/ME matching and
IR-improved DGLAP theory
in precision LHC physics scenarios. Applications to single heavy gauge
boson production at hadron colliders are illustrated.\\
\vspace{0.5cm}
\centerline{BU-HEPP-06-07, Oct., 2006, {\it presented by B.F.L. Ward at ICHEP06}}
}]
Given the approaching turn-on of the LHC, the issue of
precision predictions for the effects 
of multiple gluon and multiple photon radiative processes is 
a more immediate. A problem of some interest is the
construction of an event-by-event
precision theory of potential luminosity processes such as
single heavy gauge boson production processes.
Presuming the LHC luminosity experimental error to reach 2\%~\cite{lhclum},
the attendant theoretical precision tag on processes such as
single W,Z production should be ${\tiny\frac{2}{3}}-1\%$ so that the theoretical error
does not adversely affect the respective luminosity determination
and the attendant precision LHC physics\footnote{We note for reference that the current precision tag for the luminosity at FNAL is at the $\sim 7\%$
level~\cite{fnallum}.}. Thus, we have developed~\cite{qcedexp} a theory of
the simultaneous resummation of multiple gluon and multiple
photon radiative effects, $QED\otimes QCD$ exponentiation, to realize 
systematically the needed higher order corrections
on an event-by-event basis, in the presence of parton showers,
to the desired accuracy.
\par

We note that the new $QED\otimes QCD$ exponentiation theory
is an exact resummation theory in the spirit of the original
YFS exponentiation for $QED$ and is an
extension of the $QCD$ exponentiation theory presented
in Refs.~\cite{qcdref}. One should compare the latter to the formal proof of 
exponentiation in non-Abelian gauge theories {\it in the eikonal approximation}
as given in Ref.~\cite{gatherall}: the results in Ref.~\cite{qcdref}
are in contrast {\it exact} but have an exponent that only contains
the leading contribution of the exponent in Ref.~\cite{gatherall}-- if desired,
all of the contributions of the latter exponent can be incorporated
into the exponent Ref.~\cite{qcdref} while maintaining exactness.
Our precision requirement for the LHC luminosity processes 
requires exact ${\cal O}(\alpha_s^2,~\alpha_s\alpha,~\alpha^2)$\; $QED\otimes QCD$ exponentiation in the presence of parton showers, 
on an event-by-event basis,
realized by MC methods for realistic multiple gluon and multiple photon
radiative effects. Development of such MC's is in progress.\par
We note that in Ref.~\cite{qcedexp} we have shown that
for a process such as $q+\bar q'\rightarrow V+n(G)+n'(\gamma)+X\rightarrow
\bar{\ell} \ell'+n(g)+n'(\gamma)+X$ we have the result
{\small
\begin{equation}
\begin{split}
d\hat\sigma_{\rm exp} &= e^{\rm SUM_{IR}(QCED)}\\
   &\sum_{{n,m}=0}^\infty\int\prod_{j_1=1}^n\frac{d^3k_{j_1}}{k_{j_1}} 
\prod_{j_2=1}^m\frac{d^3{k'}_{j_2}}{{k'}_{j_2}}
\int\frac{d^4y}{(2\pi)^4}\\&e^{iy\cdot(p_1+q_1-p_2-q_2-\sum k_{j_1}-\sum {k'}_{j_2})+
D_\rQCED} \\
&\tilde{\bar\beta}_{n,m}(k_1,\ldots,k_n;k'_1,\ldots,k'_m)\frac{d^3p_2}{p_2^{\,0}}\frac{d^3q_2}{q_2^{\,0}},
\end{split}
\label{qced}
\end{equation}}\noindent
where the new YFS~\cite{yfs,yfs1} residuals, defined in Ref.~\cite{qcedexp}, 
$\tilde{\bar\beta}_{n,m}(k_1,\ldots,k_n;k'_1,\ldots,k'_m)$, with $n$ hard gluons and $m$ hard photons,
represent the successive application
of the YFS expansion first for QCD and subsequently for QED. The
functions ${\rm SUM_{IR}(QCED)},D_\rQCED$ are given in Ref.~\cite{qcedexp}.
The residuals $\tilde{\bar\beta}_{n,m}(k_1,\ldots,k_n;k'_1,\ldots,k'_m)$ 
are free of all infrared singularities
and the result in (\ref{qced}) is a representation that is exact
and that can therefore be used to make contact with parton shower 
MC's without double counting or the unnecessary averaging of effects
such as the gluon azimuthal angular distribution relative to its
parent's momentum direction.\par
   For our prototypical processes, we have the standard formula{\small
\begin{eqnarray}
d\sigma_{exp}(pp\rightarrow V+X\rightarrow \bar\ell \ell'+X') =\nonumber\\
\sum_{i,j}\int dx_idx_j F_i(x_i)F_j(x_j)d\hat\sigma_{\rm exp}(x_ix_js),
\label{sigtot} 
\end{eqnarray}}\noindent
in the notation of Ref.~\cite{qcedexp}. We will illustrate this formula with
semi-analytical methods and the case of single Z production with
the structure functions $\{F_i\}$ from Ref.~\cite{mrst1} for definiteness.
A MC realization will appear elsewhere~\cite{elsewh}.\par
As we have explained in Ref.~\cite{hep0509}, from the standpoint of using
the result (\ref{qced}) with the structure functions $\{F_i\}$, we intend
two possible approaches to the implied shower/ME matching~\cite{frixw}:
one is based on $p_T$ matching and one based on the 
shower-subtracted residuals $\hat{\tilde{\bar\beta}}_{n,m}$.
The respective shower can be that from PYTHIA~\cite{pythia}, HERWIG~\cite{herwig}, or the new shower algorithms in Ref.~\cite{jadskrz}.
This combination of showers and exact ME's can be improved systematically
to arbitrary precision order-by-order in $(\alpha_s,\alpha)$ in the presence
of exact phase space~\cite{elsewh}.\par
To illustrate the size of the interplay between QED and QCD in the threshold
region, we compute, with and without the QED, the respective ratio $r_{exp}=\sigma_{exp}/\sigma_{Born}$, with the results
{\small
\begin{equation}
r_{exp}=
\begin{cases}
1.1901&, \text{QCED}\equiv \text{QCD+QED,~~LHC}\\
1.1872&, \text{QCD,~~LHC}\\
1.1911&, \text{QCED}\equiv \text{QCD+QED,~~Tevatron}\\
1.1879&, \text{QCD,~~Tevatron.}\\
\end{cases}
\label{res1}
\end{equation}}\noindent
Here, we have used from (\ref{qced})
the $\tilde{\bar\beta}_{0,0}$-level result{\small
\begin{eqnarray}
\hat\sigma_{exp}(x_1x_2s)=\int^{v_{max}}_0 dv\gamma_{QCED} v^{\gamma_{QCED}-1}\nonumber\\
F_{YFS}(\gamma_{QCED})e^{\delta_{YFS}}\hat\sigma_{Born}((1-v)x_1x_2s)
\end{eqnarray}}
where we intend the well-known results for the 
respective parton-level Born cross
sections and the value of $v_{max}$ implied by the experimental cuts
under study. The value for the QED$\otimes$QCD
exponent is {\small
$\gamma_{QCED}= \left\{2Q_f^2\frac{\alpha}{\pi}+2C_F\frac{\alpha_s}{\pi}\right\}L_{nls}$}
where $L_{nls}=\ln x_1x_2s/\mu^2$ when $\mu$ is the factorization scale.
The functions $F_{YFS}(\gamma_{QCED})$ and $\delta_{YFS}(\gamma_{QCED})$
are well-known~\cite{yfs1} as well:{\small
\begin{equation}
F_{YFS}(\gamma_{QCED})=\frac{e^{-\gamma_{QCED}\gamma_E}}{\Gamma(1+\gamma_{QCED})},
\label{yfsfn}
\end{equation}}
with
$\delta_{YFS}(\gamma_{QCED})=\frac{1}{4}\gamma_{QCED}
+(Q_f^2\frac{\alpha}{\pi}+C_F\frac{\alpha_s}{\pi})(2\zeta(2)-\frac{1}{2})$,
where $\zeta(2)$ is Riemann's zeta function of argument 2, i.e., $\pi^2/6$,
and $\gamma_E$ is Euler's constant, i.e., 0.5772... .\par

We see that the size of the QED effects, $\sim 0.3\%$, is appreciable in
a $1\%$ precision tag budget and 
these effects are comparable to the QED effects
found for the structure function evolution itself in Refs.~\cite{cern2000,spies,james1,roth,james2}. We also see that our soft, 
QCD correction is entirely consistent with the exact results in Ref.~\cite{van1,van2,anas} and that our QED effects are consistent with the exact ${\cal O}(\alpha)$ results in Refs.~\cite{baurall,ditt,russ}. Such cross-checks are essential
in establishing a rigorous precision tag on the theoretical predictions.\par
\begin{figure}
\begin{center}
\epsfig{file=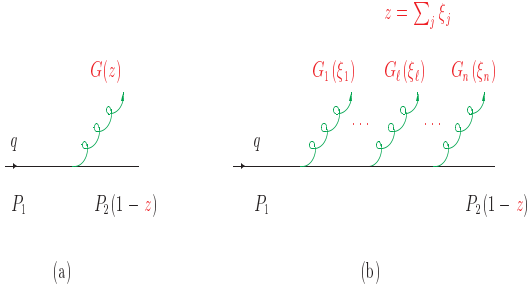,width=70mm,height=38mm}
\end{center}
\caption{\baselineskip=7mm  In (a), we show the usual process $q\rightarrow q(1-z)+G(z)$;
in (b), we show its multiple gluon improvement $q\rightarrow q(1-z)+G_1(\xi_1)+\cdots+G_n(\xi_n),~~z=\sum_j\xi_j$.}
\label{fig1-a}
\end{figure}

With an eye toward the 1\% precision tag for the single heavy gauge 
boson production
at the LHC, we have analyzed~\cite{irdglap} the IR limit of the DGLAP~\cite{dglap} kernels themselves. The effects which we address are illustrated in Fig.~\ref{fig1-a}.
We apply the QCD exponentiation master formula from Ref.~\cite{qcdref},
the analog of (\ref{qced}) for just QCD, 
to the gluon emission transition that
corresponds to $P_{qq}(z)$, i.e., to the squared amplitude for
$q\rightarrow q(z)+G(1-z)$; as we show in Ref.~\cite{irdglap}, this allows us toget the following IR-improved DGLAP kernels
{\small
\begin{equation}
\begin{split}
P_{qq}^{exp}(z)&= C_F F_{YFS}(\gamma_q)e^{\frac{1}{2}\delta_q}\{\frac{1+z^2}{1-z}(1-z)^{\gamma_q}\\
& -f_q(\gamma_q)\delta(1-z)\},
\end{split}
\end{equation}
\begin{equation}
P_{Gq}(z)^{exp} = C_F F_{YFS}(\gamma_q)e^{\frac{1}{2}\delta_q}\frac{1+(1-z)^2}{z} z^{\gamma_q},
\end{equation}
\begin{align}
P_{GG}^{exp}(z)&= 2C_G F_{YFS}(\gamma_G)e^{\frac{1}{2}\delta_G}\{ \frac{1-z}{z}z^{\gamma_G}\nonumber\\
&+\frac{z}{1-z}(1-z)^{\gamma_G}
+\frac{1}{2}(z^{1+\gamma_G}(1-z)\nonumber\\
&+z(1-z)^{1+\gamma_G})- f_G(\gamma_G) \delta(1-z)\},\\
P_{qG}^{exp}(z)&= F_{YFS}(\gamma_G)e^{\frac{1}{2}\delta_G}\frac{1}{2}\{ z^2(1-z)^{\gamma_G}\nonumber\\
&+(1-z)^2z^{\gamma_G}\},
\label{dglap22}
\end{align}}
where from the standard DGLAP methodology~\cite{irdglap} we have~{\small
$f_q(\gamma_q)=\frac{2}{\gamma_q}-\frac{2}{\gamma_q+1}+\frac{1}{\gamma_q+2}$,
$\gamma_q = C_F\frac{\alpha_s}{\pi}t=\frac{4C_F}{\beta_0}$,
$\delta_q=\frac{\gamma_q}{2}+\frac{\alpha_sC_F}{\pi}(\frac{\pi^2}{3}-\frac{1}{2})$,
$\gamma_G = C_G\frac{\alpha_s}{\pi}t=\frac{4C_G}{\beta_0}$,
$\delta_G =\frac{\gamma_G}{2}+\frac{\alpha_sC_G}{\pi}(\frac{\pi^2}{3}-\frac{1}{2})$}, and {\small
\begin{align}
f_G(\gamma_G)&=\frac{n_f}{C_G}\frac{1}{(1+\gamma_G)(2+\gamma_G)(3+\gamma_G)}\nonumber\\
&+\frac{2}{\gamma_G(1+\gamma_G)(2+\gamma_G)}\nonumber\\
&+\frac{1}{(1+\gamma_G)(2+\gamma_G)}
+\frac{1}{2(3+\gamma_G)(4+\gamma_G)}\nonumber\\
&+\frac{1}{(2+\gamma_G)(3+\gamma_G)(4+\gamma_G)}.
\label{dglap19}
\end{align}}
Here, $\beta_0=11-\frac{2}{3}n_f$, where $n_f$ is the number of
active quark flavors.
We show in Ref~\cite{irdglap,hep-ph0602025} that these improved kernels 
change significantly the evolution of the moments of the structure functions.
For example, the non-singlet anomalous dimension $A^{NS}_n$ becomes,
from the exponentiated kernels, 
{\small
\begin{align}
A^{NS}_n&= C_F F_{YFS}(\gamma_q)e^{\frac{1}{2}\delta_q}\nonumber\\
&[B(n,\gamma_q)+B(n+2,\gamma_q)-f_q(\gamma_q)]
\label{dglap25}
\end{align}}
where $B(x,y)$ is the beta function
given by {\small$B(x,y)=\Gamma(x)\Gamma(y)/\Gamma(x+y)$.}
Compare the usual result{\small
\begin{equation}
A^{NS^o}_n\equiv C_F [-\frac{1}{2}+\frac{1}{n(n+1)}-2\sum_{j=2}^{n}\frac{1}{j}].
\label{dglap26}
\end{equation}}
The IR-improved n-th moment goes for large $n$ to a multiple of $-f_q$,
consistent with $\lim_{n\rightarrow \infty}z^{n-1} = 0$ for $0\le z<1$;
the usual result diverges as $-2C_F\ln n$.
The two results differ for finite $n$ as well: 
we get, for example, for $\alpha_s\cong .118$,
$A^{NS}_2 =C_F(-1.33),C_F(-0.966)$ for (\ref{dglap26}) and (\ref{dglap25}), respectively. See Refs.~\cite{irdglap,hep-ph0602025} for further discussion.
\par
Contact with the exact 2-loop and 3-loop 
results in Refs.~\cite{mvermn,mvovermn}
has also been made~\cite{irdglap}: for the non-singlet case, we have 
, in the notation of Ref.~\cite{mvovermn}, 
the exact 3-loop IR-improved result{\small
\begin{equation}
\begin{split}
P_{ns}^{+,exp}(z)=(\frac{\alpha_s}{4\pi}) 2P_{qq}^{exp}(z)+F_{YFS}(\gamma_q)e^{\frac{1}{2}\delta_q}\\
\big{[}(\frac{\alpha_s}{4\pi})^2\{(1-z)^{\gamma_q}\bar{P}_{ns}^{(1)+}(z)+\bar{B}_2\delta(1-z)\}\\
+(\frac{\alpha_s}{4\pi})^3\{(1-z)^{\gamma_q}\bar{P}_{ns}^{(2)+}(z)+\bar{B}_3\delta(1-z)\}\big{]}
\end{split}
\label{vermn3}
\end{equation}} 
where $P_{qq}^{exp}(z)$ is given in (\ref{dglap22}) and the resummed residuals 
$\bar{P}_{ns}^{(i)+}$,~$i=1,2$ are related to the exact results~\cite{mvermn,mvovermn} for $P_{ns}^{(i)+}$,~$i=1,2$, via{\small
\begin{equation}
\bar{P}_{ns}^{(i)+}(z)=P_{ns}^{(i)+}(z)-B_{1+i}\delta(1-z)
+\Delta_{ns}^{(i)+}(z),
\label{vermn4}
\end{equation}}
where the $B_i$~\cite{mvermn,mvovermn} and the $\Delta_{ns}^{(i)+}(z)$ are 
given explicitly in Ref.~\cite{irdglap}.
The detailed phenomenological consequences of the fully exponentiated
2- and 3-loop DGLAP kernel set will appear elsewhere~\cite{elsewh}.
\par
In summary, the methods illustrated and presented herein are all under 
investigation and implementation as we prepare for the era of 1\% precision
LHC theoretical predictions on the necessary processes. Finally,
one of us ( B.F.L.W.)
thanks Prof. W. Hollik for the support and kind
hospitality of the MPI, Munich, while a part of this work was
completed. We also thank Prof. S. Jadach for useful discussions. This work is partly supported by US DOE grant DE-FG02-05ER41399 and by NATO grant PST.CLG.980342.

\end{document}